\documentstyle[amssymb,aps]{revtex}

\begin{document}
\title{Return operator, cross Bell bases and protocol of teleportation of arbitrary
multipartite qubit entanglement}
\author{Zai-Zhe Zhong}
\address{Department of Physics, Liaoning Normal University, Dalian 116029, Liaoning,\\
China. E-mail: zhongzaizheh@hotmail.com}
\maketitle

\begin{abstract}
In this paper, we define the return operator, the cross product operator and
the cross Bell bases. Using the cross Bell bases, we give a protocol of
teleportation of arbitrary multipartite qubit entanglement, this scheme is a
quite natural generalization of the BBCJPW scheme. We find that this
teleportation, in fact, is essentially determined by the teleportation of
every single unknown qubit state as in the original scheme of BBCJPW. The
calculation in detail is given for the case of tripartite qubit.

PACC numbers: 03.67.Mn, 03.65.Ud, 03.67.Hk.
\end{abstract}

The quantum teleportation is a quite interesting and important topic in
modern quantum mechanics and quantum information (about this topic and
references, can see [1]). The original scheme of BBCJPW[2] is to discuss the
teleportation of single unknown qubit pure-state, in which there is no
teleportation problem of quantum entanglement of course. In view of
application of quantum information theory, it obviously is not enough,, we
must consider the problems of teleportation of multipartite quantum
entangled states, first the multipartite qubit entangled states. Recently,
there are some works (e.g. see [3-6]) discussing the problems of
teleportation of arbitrary multipartite qubit entanglement, especially the
bipartite and tripartite cases. However, in these schemes, generally, the
mathematical form and concrete operation in physics seem more complex, the
relation between them and BBCJPW scheme is not quite clear, we must find
more perfect scheme. We find that the key of problems is the order of
factors in the tensor products of the entangled state taken as channel. In
this paper we give a new way to discuss the problem of teleportation of
arbitrary multipartite qubit entanglement. First, in order to overcome the
puzzle brought by order problem, we define a `return operator' and the
`cross product' operator. Next we define what is the cross Bell basis. Using
the cross Bell bases, we give a improved protocol of teleportation of
arbitrary multipartite qubit entanglement, it is a very natural
generalization of the original scheme of BBCJPW, the latter only is an
extremity case (N=1) of the former. We find that this teleportation, in
fact, is essentially determined by the teleportation of every single unknown
qubit state as in the original scheme of BBCJPW[$2].$ In mathematical form
our scheme also very simple. For the case of tripartite qubit, we make the
concrete calculation.

The train of thought in this paper is as follows. In the following we write
the Bell states of particles 1 and 2 as 
\begin{equation}
\mid \phi _{12}^{\pm }\rangle =\frac 1{\sqrt{2}}\left( \mid 0_10_2\rangle
\pm \mid 1_11_2\rangle \right) ,\mid \psi _{12}^{\pm }\rangle =\frac 1{\sqrt{%
2}}\left( \mid 0_11_2\rangle \pm \mid 1_10_2\rangle \right)
\end{equation}
Now we look back the character of standard teleportation BBCJPW scheme[2].
Suppose that Alice holds a unknown state $\mid \varphi _3\rangle =\alpha
\mid 0_3\rangle +\beta \mid 1_3\rangle \left( \left| \alpha \right|
^2+\left| \beta \right| ^2=1\right) ,$ two particles 1, 2 are in Alice and
Bob respectively, particles 1, 2 are in a Bell state (quantum channel), say $%
\mid \psi _{12}^{-}\rangle .$ It is known that the realization of the
standard teleportation, in fact, completely dependents on the decomposition
of the total state $\mid \Psi _{total}\rangle =\mid \psi _{12}^{-}\rangle
\mid \varphi _3\rangle $ as 
\begin{eqnarray}
&\mid &\Psi _{total}\rangle =\mid \psi _{12}^{-}\rangle \mid \varphi
_3\rangle =\frac 1{\sqrt{2}}\left( \mid 0_11_2\rangle -\mid 1_10_2\rangle
\right) \left( \alpha \mid 0_3\rangle +\beta \mid 1_3\rangle \right) 
\nonumber \\
&=&\frac 1{\sqrt{2}}\left\{ \mid 0_1\rangle \left( \alpha \mid 1_20_3\rangle
+\beta \mid 1_21_3\rangle \right) -\mid 1_1\rangle \left( \alpha \mid
0_20_3\rangle +\beta \mid 0_21_3\rangle \right) \right\}  \nonumber \\
&=&\frac 1{\sqrt{2}}\left\{ 
\begin{array}{c}
\mid 0_1\rangle \left[ -\alpha \frac 1{\sqrt{2}}\left( \mid \psi
_{23}^{+}\rangle -\mid \psi _{23}^{-}\rangle \right) -\beta \frac 1{\sqrt{2}}%
\left( \mid \phi _{23}^{+}\rangle -\mid \phi _{23}^{-}\rangle \right) \right]
\\ 
-\mid 1_1\rangle \left[ \alpha \frac 1{\sqrt{2}}\left( \mid \phi
_{23}^{+}\rangle +\mid \phi _{23}^{-}\rangle \right) -\beta \frac 1{\sqrt{2}}%
\left( \mid \psi _{23}^{+}\rangle +\mid \psi _{23}^{-}\rangle \right) \right]
\end{array}
\right\} \\
&=&\frac 12\left\{ 
\begin{array}{c}
\left( -\beta \mid 0_1\rangle -\alpha \mid 1_1\rangle \right) \mid \phi
_{23}^{+}\rangle +\left( \beta \mid 0_1\rangle -\alpha \mid 1_1\rangle
\right) \mid \phi _{23}^{-}\rangle \\ 
+\left( -\alpha \mid 0_1\rangle +\beta \mid 1_1\rangle \right) \mid \psi
_{23}^{+}\rangle +\left( \alpha \mid 0_1\rangle +\beta \mid 1_1\rangle
\right) \mid \psi _{23}^{-}\rangle
\end{array}
\right\}  \nonumber
\end{eqnarray}
The rest steps are standard, i.e. to take a Bell measurement for particles
2, 3 by Alice, and the wave function will collapse with probability $\frac 14
$ to one of $\left( -\beta \mid 0_1\rangle -\alpha \mid 1_1\rangle \right)
\mid \phi _{23}^{+}\rangle $, $\cdots ,$ $\left( \alpha \mid 0_1\rangle
+\beta \mid 1_1\rangle \right) \mid \psi _{23}^{-}\rangle ,$ etc. We see
that the essence of this scheme are to use the decomposition of a product
state$\mid ij\rangle $ by the Bell bases and the Bell measurement for
particles 2, 3. We can figure it as follows 
\begin{equation}
\stackrel{\mid \varphi _3\rangle }{\bigcirc _{3,Alice}}\;\otimes \;\stackrel{%
\mid \psi _{23}^{-}\rangle }{\bigcirc _{2,Alice}====\bigcirc _{1,\;Bob}}%
\stackrel{
\begin{array}{c}
\text{Bell\ measurement } \\ 
\text{for\ 2,3}
\end{array}
}{\longrightarrow \longrightarrow \longrightarrow \longrightarrow }\stackrel{
\begin{array}{c}
\text{collapse\ into\ one} \\ 
\text{Bell\ state}
\end{array}
}{\bigcirc _{3,Alice}====\bigcirc _{2,Alice}}\;\otimes \;\stackrel{\text{%
result}}{\bigcirc _{1,Bob}}
\end{equation}
where ==== denotes a Bell state. In this paper, the figure of our scheme of
teleportation of a N-partite $\left( N\geqslant 2\right) $ qubit entangled
state $\mid \varphi _3^{\left( n\right) }\rangle =\sum_{i_1,\cdots
,i_n=0,1}\alpha _{i_1\cdots i_n}\mid i_1\cdots i_n\rangle $ $\left(
\sum_{i_1,\cdots ,i_n=0,1}\left| \alpha _{i_1\cdots i_n}\right| ^2=1\right) $
is as follows, which obviously is the generalization of the above figure 
\begin{eqnarray}
&& 
\begin{array}{lll}
\left( 
\begin{array}{l}
\text{Entangled state }\mid \varphi ^{\left( N\right) }\rangle ,\text{ Alice}
\\ 
\bigcirc _{2N+1} \\ 
\bigcirc _{2N+N} \\ 
\vdots \\ 
\bigcirc _{3N}
\end{array}
\right) & \otimes & \left( 
\begin{array}{lll}
\text{Alice} &  & \text{Bob} \\ 
\bigcirc _{N+1} & ==== & \bigcirc _1 \\ 
\bigcirc _{N+2} & ==== & \bigcirc _2 \\ 
\vdots & \vdots & \vdots \\ 
\bigcirc _{2N} & ==== & \bigcirc _N
\end{array}
\right)
\end{array}
\nonumber \\
&&\stackrel{
\begin{array}{c}
\text{Bell\ measurements for pairs } \\ 
\left( N+1,2N+1\right) ,\cdots ,\left( 2N,3N\right)
\end{array}
}{\longrightarrow \longrightarrow \longrightarrow \longrightarrow
\longrightarrow \longrightarrow } 
\begin{array}{ll}
\text{collapses\ into\ one of} & \left( 
\begin{array}{lll}
\text{Alice} &  & \text{Alice} \\ 
\bigcirc _{N+1} & ==== & \bigcirc _{2N+1} \\ 
\bigcirc _{N+2} & ==== & \bigcirc _{2N+2} \\ 
\vdots & \vdots & \vdots \\ 
\bigcirc _{2N} & ==== & \bigcirc _{3N}
\end{array}
\right)
\end{array}
\otimes \left( 
\begin{array}{l}
\text{result, Bob} \\ 
\bigcirc _1 \\ 
\bigcirc _2 \\ 
\vdots \\ 
\bigcirc _N
\end{array}
\right)
\end{eqnarray}
where $\left( 
\begin{array}{l}
==== \\ 
==== \\ 
\vdots \\ 
====
\end{array}
\right) $ denotes some `cross Bell bases'\ (see below).

The concrete steps are as follows. In the first place, we need to define the
return operator and the cross product operator $\nabla .$

{\it Definition 1. }The so-called `return operator' ${\cal R}$ is a linear
operator, its action is to return the factor order of every term (a tensor
product of single qubit state) of a N-partite qubit pure-state $\mid \Psi
\rangle $ into the natural order as $\mid i_1i_2\cdots i_N\rangle .$

{\bf Example.} ${\cal R}\left( a\mid 0_20_10_3\cdots 0_N\rangle +b\mid
1_30_21_10\cdots 0_N\rangle \right) =a\mid 0_10_20_3\cdots 0_N\rangle +b\mid
1_10_21_30\cdots 0_N\rangle .$

Obviously the operator ${\cal R}$ does not break the normalization and
orthogonality, i.e. if $\mid \Psi \rangle $ is normal, then ${\cal R}\left(
\mid \Psi \rangle \right) $ is also, and If $\langle \Phi \mid \Psi \rangle
=0$, then $\langle {\cal R}\left( \Phi \right) \mid {\cal R}\left( \mid \Psi
\rangle \right) \rangle =0.$

{\it Definition 2. }If $\left\{ \mid \Psi _{\left( n\right) }\rangle
\right\} \left( n=1,\cdots ,N\right) $ is a sequence of N qubit pure-states $%
\mid \Psi _{\left( n\right) }\rangle
=\sum\limits_{i_n,j_n=0,1}c_{i_nj_n}\mid i_nj_n\rangle ,$ the cross product
of (ordered) finite sequence $\left\{ \mid \Psi _{\left( n\right) }\rangle
\right\} $ written by notation $\nabla \left( \mid \Psi _{\left( 1\right)
}\rangle ,\cdots ,\mid \Psi _{\left( n\right) }\rangle \right) ,$ which is a
2N-partite qubit pure-state, is defined to be the result ${\cal R}\left(
\otimes _{n=1}^N\mid \Psi _{\left( n\right) }\rangle \right) $of the
ordinary tensor product $\otimes _{n=1}^N\mid \Psi _{\left( n\right)
}\rangle =\sum_{i_1,\cdots ,i_N}\left( \prod_{n=1}^Nc_{i_nj_n}\right) \mid
i_1j_1\cdots i_Nj_N\rangle ,$ where ${\cal R}$ is the return operator with
respect to the order $\left( i_1,\cdots ,i_N,j_1,\cdots ,j_N\right) $, i.e. 
\begin{equation}
\nabla \left( \mid \Psi _{\left( 1\right) }\rangle ,\cdots ,\mid \Psi
_{\left( n\right) }\rangle \right) =\sum_{i_1,\cdots ,i_N}\left(
\prod_{n=1}^Nc_{i_nj_n}\right) \mid i_1\cdots i_Nj_1\cdots j_N\rangle
\end{equation}

We notice that although $\nabla \left( \mid \Psi _{\left( 1\right) }\rangle
,\cdots ,\mid \Psi _{\left( n\right) }\rangle \right) $, in fact, has the
completely same physical contents to $\otimes _{n=1}^N\mid \Psi _{\left(
n\right) }\rangle ,$ in our scheme the expression $\nabla \left( \mid \Psi
_{\left( 1\right) }\rangle ,\cdots ,\mid \Psi _{\left( n\right) }\rangle
\right) $ for the order $\left( i_1,\cdots ,i_N,j_1,\cdots ,j_N\right) $ is
very important, it is just the reason to introduce the cross product
operator $\nabla $ in this paper$.$ Obviously, the operator $\nabla $ is
linear for every $\mid \Psi _{\left( i\right) }\rangle ,$ e.g. $\nabla
\left( \alpha \mid \Psi _{\left( 1\right) }\rangle +\beta \mid \Psi _{\left(
1\right) }^{\prime }\rangle ,\mid \Psi _{\left( 2\right) }\rangle \cdots
,\mid \Psi _{\left( n\right) }\rangle \right) =\alpha \nabla \left( \mid
\Psi _{\left( 1\right) }\rangle ,\mid \Psi _{\left( 2\right) }\rangle
,\cdots ,\mid \Psi _{\left( n\right) }\rangle \right) +$ $\beta \nabla
\left( \mid \Psi _{\left( 1\right) }^{\prime }\rangle ,\mid \Psi _{\left(
2\right) }\rangle ,\cdots ,\mid \Psi _{\left( n\right) }\rangle \right)
,\cdots ,$ etc.

Now by use of cross products, we can construct a basis of the Hilbert space $%
H_1\otimes H_2\otimes \cdots \otimes H_{2N},$ where every $H_r\left(
r=1,\cdots ,2N\right) $ is a Hilbert space of qubit states. For the sake of
convenience, in the following we simply write the Bell bases as 
\begin{equation}
\mid \Omega _{1,2}^{\left( 1\right) }\rangle \equiv \mid \phi
_{12}^{+}\rangle ,\mid \Omega _{1,2}^{\left( 2\right) }\rangle \equiv \mid
\phi _{12}^{-}\rangle ,\mid \Omega _{1,2}^{\left( 3\right) }\rangle \equiv
\mid \psi _{12}^{+}\rangle ,\mid \Omega _{1,2}^{\left( 4\right) }\rangle
\equiv \mid \psi _{12}^{-}\rangle
\end{equation}

{\it Definition 3.} Let the set ${\Bbb B}_{1,\cdots ,2N}$ be defined by 
\begin{eqnarray}
{\Bbb B}_{1,\cdots ,2N} &=&\left\{ {\Bbb B}_{1,\cdots ,2N}^{\left( \lambda
_1,\lambda _2,\cdots \lambda _N\right) }\text{ for all possible }\lambda
_1,\lambda _2,\cdots ,\lambda _N=1,2,3,4\right\} \text{ }  \nonumber \\
{\Bbb B}_{1,\cdots ,2N}^{\left( \lambda _1,\lambda _2,\cdots \lambda
_N\right) } &\equiv &\nabla \left( \mid \Omega _{1,n+1}^{\left( \lambda
_1\right) }\rangle ,\mid \Omega _{2,n+2}^{\left( \lambda _2\right) }\rangle
,\cdots ,\mid \Omega _{N,2N}^{\left( \lambda _N\right) }\rangle \right) 
\text{ }
\end{eqnarray}
It is easily verified that ${\Bbb B}_{1,\cdots ,2N}$ has 2$^{4N}$ entries
and surly forms a complete orthogonal basis of $H_1\otimes H_2\otimes \cdots
\otimes H_{2N}$, we call it the `cross Bell basis', which is a
generalization of the ordinary Bell bases. Here, it must be stressed that it
is different from the ordinary Bell bases that {\bf all cross Bell bases for
any N}$\geqslant 3${\bf \ are not maximal entangled states}. In fact, they
obviously are entangled, but partially separable[7,8] states, e.g., all $%
{\Bbb B}_{1234}^{\left( r,s\right) }$ are 13-24 entangled, etc. In the
following, we shall take one entry of ${\Bbb B}_{1,\cdots ,2N}$ as the
quantum channel, this fact also identifies just with the view of [9].

These bases will construct the parts described by ==== as the figure as in
Eq.($4$). By use of the following equations 
\begin{eqnarray}
&\mid &0_10_2\rangle =\frac 1{\sqrt{2}}\left( \mid \phi _{12}^{+}\rangle
+\mid \phi _{12}^{-}\rangle \right) ,\mid 1_11_2\rangle =\frac 1{\sqrt{2}}%
\left( \mid \phi _{12}^{+}\rangle -\mid \phi _{12}^{-}\rangle \right) 
\nonumber \\
&\mid &0_11_2\rangle =\frac 1{\sqrt{2}}\left( \mid \psi _{12}^{+}\rangle
+\mid \psi _{12}^{-}\rangle \right) ,\mid 1_10_2\rangle =\frac 1{\sqrt{2}}%
\left( \mid \psi _{12}^{+}\rangle -\mid \psi _{12}^{-}\rangle \right)
\end{eqnarray}
we obtain the transformation relation between the natural basis and cross
Bell basis as follows 
\begin{eqnarray}
&\mid &0_10_2\cdots 0_{2N-1}0_{2N}\rangle =2^{-\frac N2}\nabla \left( \mid
\Omega _{1,N+1}^{\left( 1\right) }\rangle +\mid \Omega _{1,N+1}^{\left(
2\right) }\rangle ,\cdots ,\mid \Omega _{N,2N}^{\left( 1\right) }\rangle
+\mid \Omega _{N,2N}^{\left( 2\right) }\rangle \right)  \nonumber \\
&=&2^{-\frac N2}\sum_{\lambda _1,\lambda _2,\cdots ,\lambda _N=1,2,}{\Bbb B}%
_{1,\cdots ,2N}^{\left( \lambda _1,\lambda _2,\cdots \lambda _N\right) } 
\nonumber \\
&\mid &0_10_2\cdots 0_{2N-1}1_{2N}\rangle =2^{-\frac N2}\nabla \left( \mid
\Omega _{1,N+1}^{\left( 1\right) }\rangle +\mid \Omega _{1,N+1}^{\left(
2\right) }\rangle ,\cdots ,\mid \Omega _{N-1,2N-1}^{\left( 1\right) }\rangle
+\mid \Omega _{N-1,2N-1}^{\left( 2\right) }\rangle ,\mid \Omega
_{N,2N}^{\left( 3\right) }\rangle +\mid \Omega _{N,2N}^{\left( 4\right)
}\rangle \right)  \nonumber \\
&=&2^{-\frac N2}\sum_{\lambda _1,\lambda _2,\cdots ,=1,2.\text{ }\lambda
_N=3,4}{\Bbb B}_{1,\cdots ,2N}^{\left( \lambda _1,\lambda _2,\cdots \lambda
_N\right) }  \nonumber \\
&\mid &0_10_2\cdots 1_{2N-1}0_{2N}\rangle =2^{-\frac N2}\nabla \left( 
\begin{array}{c}
\mid \Omega _{1,N+1}^{\left( 1\right) }\rangle +\mid \Omega _{1,N+1}^{\left(
2\right) }\rangle ,\cdots ,\mid \Omega _{N-2,2N-2}^{\left( 1\right) }\rangle
+\mid \Omega _{N-2,2N-2}^{\left( 2\right) }\rangle , \\ 
\mid \Omega _{N-1,2N-1}^{\left( 3\right) }\rangle -\mid \Omega
_{N-1,2N-1}^{\left( 4\right) }\rangle ,\mid \Omega _{N,2N}^{\left( 1\right)
}\rangle +\mid \Omega _{N,2N}^{\left( 2\right) }\rangle
\end{array}
\right)  \nonumber \\
&=&\left( -1\right) ^{\lambda _{N-1}+1}2^{-\frac N2}\sum_{\lambda _1,\lambda
_2,\cdots ,\lambda _{N-2},\lambda _N=1,2.\;\lambda _{N-1}=3,4}{\Bbb B}%
_{1,\cdots ,2N}^{\left( \lambda _1,\lambda _2,\cdots \lambda _N\right) } 
\nonumber \\
&&\cdots \cdots \cdots \cdots \\
&\mid &1_11_2\cdots 1_{2N-1}0_{2N}\rangle =2^{-\frac N2}\nabla \left( \mid
\Omega _{1,N+1}^{\left( 1\right) }\rangle -\mid \Omega _{1,N+1}^{\left(
2\right) }\rangle ,\cdots ,\mid \Omega _{N-1,2N-11}^{\left( 1\right)
}\rangle -\mid \Omega _{N-1,2N+1}^{\left( 2\right) }\rangle ,\mid \Omega
_{N,2N}^{\left( 3\right) }\rangle -\mid \Omega _{N,2N}^{\left( 4\right)
}\rangle \right)  \nonumber \\
&=&\left( -1\right) ^{\sum_{s=1}^N\lambda _s-N}2^{-\frac N2}\sum_{\lambda
_1,\lambda _2,\cdots ,=1,2.\text{ R}_N=3,4}{\Bbb B}_{1,\cdots ,2N}^{\left(
\lambda _1,\lambda _2,\cdots \lambda _N\right) }  \nonumber \\
&\mid &1_11_2\cdots 1_{2N-1}1_{2N}\rangle =2^{-\frac N2}\nabla \left( \mid
\Omega _{1,N+1}^{\left( 1\right) }\rangle -\mid \Omega _{1,N+1}^{\left(
2\right) }\rangle ,\cdots ,\mid \Omega _{N,2N}^{\left( 1\right) }\rangle
-\mid \Omega _{N,2N}^{\left( 2\right) }\rangle \right)  \nonumber \\
&=&\left( -1\right) ^{\sum_{s=1}^N\lambda _s-N}2^{-\frac N2}\sum_{\lambda
_1,\lambda _2,\cdots ,\lambda _N=1,2}{\Bbb B}_{1,\cdots ,2N}^{\left( \lambda
_1,\lambda _2,\cdots \lambda _N\right) }  \nonumber
\end{eqnarray}
We still need to use the following common relations 
\begin{eqnarray}
&\mid &\phi _{12}^{\pm }\rangle \mid 0_3\rangle =\frac 12\left[ \mid
0_1\rangle \left( \mid \phi _{23}^{+}\rangle +\mid \phi _{23}^{-}\rangle
\right) \pm \mid 1_1\rangle \left( \mid \psi _{23}^{+}\rangle -\mid \psi
_{23}^{-}\rangle \right) \right]  \nonumber \\
&\mid &\phi _{12}^{\pm }\rangle \mid 1_3\rangle =\frac 12\left[ \mid
0_1\rangle \left( \mid \psi _{23}^{+}\rangle +\mid \psi _{23}^{-}\rangle
\right) \pm \mid 1_1\rangle \left( \mid \phi _{23}^{+}\rangle -\mid \phi
_{23}^{-}\rangle \right) \right]  \nonumber \\
&\mid &\phi _{12}^{\pm }\rangle \mid 0_3\rangle =\frac 12\left[ \mid
0_1\rangle \left( \mid \psi _{23}^{+}\rangle -\mid \psi _{23}^{-}\rangle
\right) \pm \mid 1_1\rangle \left( \mid \phi _{23}^{+}\rangle +\mid \phi
_{23}^{-}\rangle \right) \right] \\
&\mid &\phi _{12}^{\pm }\rangle \mid 1_3\rangle =\frac 12\left[ \mid
0_1\rangle \left( \mid \phi _{23}^{+}\rangle -\mid \phi _{23}^{-}\rangle
\right) \pm \mid 1_1\rangle \left( \mid \psi _{23}^{+}\rangle +\mid \psi
_{23}^{-}\rangle \right) \right]  \nonumber
\end{eqnarray}
They can be simply reduced as $\left( i=0,1\right) $%
\begin{eqnarray}
&\mid &\Omega _{12}^{\left( \lambda \right) }\rangle \mid i_3\rangle =\frac 1%
2\left[ 
\begin{array}{c}
\mid 0_1\rangle \left( \mid \Omega _{23}^{\left( 1+2i\right) }\rangle +\mid
\Omega _{23}^{\left( 2+2i\right) }\rangle \right) \\ 
+\left( -1\right) ^{\lambda +1}\mid 1_1\rangle \left( \mid \Omega
_{23}^{\left( 3-2i\right) }\rangle -\mid \Omega _{23}^{\left( 4-2i\right)
}\rangle \right)
\end{array}
\right] \text{ for }\lambda =1,2  \nonumber \\
&\mid &\Omega _{12}^{\left( \lambda \right) }\rangle \mid i_3\rangle =\frac 1%
2\left[ 
\begin{array}{c}
\mid 0_1\rangle \left( \mid \Omega _{23}^{\left( 3+2i\;%
\mathop{\rm mod}%
4\right) }\rangle -\mid \Omega _{23}^{\left( 4+2i\;%
\mathop{\rm mod}%
4\right) }\rangle \right) + \\ 
\left( -1\right) ^{\lambda +1}\mid 1_1\rangle \left( \mid \Omega
_{23}^{\left( 1+2i\;%
\mathop{\rm mod}%
4\right) }\rangle +\mid \Omega _{23}^{\left( 2+2i\;%
\mathop{\rm mod}%
4\right) }\rangle \right)
\end{array}
\right] \text{ for }\lambda =3,4
\end{eqnarray}

Now, by use of the above cross Bell bases and relations, we can realize the
teleportation of an unknown N-partite qubit entangled state. For the sake of
simplicity, we only concretely discuss the case of tripartite qubit states.
As for the simplest case of N=2, the final result is same to [3], however in
the latter the way is more complex; For higher dimensional cases, it is
similar, only the calculation becomes more complex. We take a cross Bell
basic entry, say ${\Bbb B}_{1,2,3,4,5,6}^{\left( 4,4,4\right) }=\nabla
\left( \psi _{14}^{-},\psi _{25}^{-},\psi _{36}^{-}\right) ,$ as the quantum
channel, i.e. particles 1,2,3 are in Bob, particles 4,5,6 are in Alice and
1,2,3,4,5,6 are in the entangled state ${\Bbb B}_{1,2,3,4,5,6}^{\left(
4,4,4\right) }=\nabla \left( \psi _{14}^{-},\psi _{25}^{-},\psi
_{36}^{-}\right) .$ Suppose that Alice holds particles 7,8,9 which are in an
unknown entangled state as $\mid \varphi _{789}\rangle
=\sum_{i_7,i_8,i_9=0,1}\alpha _{_{i_7i_8i_9}}\mid i_7i_8i_9\rangle ,$ where $%
\sum_{i_7,i_8,i_9=0,1}\left| \alpha _{_{i_7i_8i_9}}\right| ^2=1.$ The total
wave function is 
\begin{eqnarray}
&\mid &\Psi _{123456789}\rangle ={\Bbb B}_{1,2,3,4,5,6}^{\left( 4,4,4\right)
}\otimes \mid \varphi _{789}\rangle ={\cal R}\left( \psi _{14}^{-}\otimes
\psi _{25}^{-}\otimes \psi _{36}^{-}\otimes \mid \varphi _{789}\rangle
\right)  \nonumber \\
&=&\sum_{i_7,i_8,i_9=0,1}\alpha _{_{i_7i_8i_9}}{\cal R}\left( \left[ \psi
_{14}^{-}\otimes \mid i_7\rangle \right] \otimes \left[ \psi
_{25}^{-}\otimes \mid i_8\rangle \right] \otimes \left[ \psi
_{36}^{-}\otimes \mid i_9\rangle \right] \right)
\end{eqnarray}
where ${\cal R}$ denotes the above return operation with respect to the
natural order $\left( 1,2,\cdots ,9\right) .$ Using Eqs.(8-11), we obtain 
\begin{eqnarray}
&\mid &\Psi _{123456789}\rangle =\frac 18\left\{ \alpha _{0_70_80_9}{\cal R}%
\left[ 
\begin{array}{c}
\mid 0_1\rangle \left( \psi _{47}^{+}-\psi _{47}^{-}\right) -\mid 1_1\rangle
\left( \phi _{47}^{+}+\phi _{47}^{-}\right) \\ 
\otimes \mid 0_1\rangle \left( \psi _{58}^{+}-\psi _{58}^{-}\right) -\mid
1_1\rangle \left( \phi _{58}^{+}+\phi _{58}^{-}\right) \\ 
\otimes \mid 0_1\rangle \left( \psi _{69}^{+}-\psi _{69}^{-}\right) -\mid
1_1\rangle \left( \phi _{69}^{+}+\phi _{69}^{-}\right)
\end{array}
\right] \right\}  \nonumber \\
&&+\alpha _{0_70_81_9}{\cal R}\left[ 
\begin{array}{c}
\mid 0_1\rangle \left( \psi _{47}^{+}-\psi _{47}^{-}\right) -\mid 1_1\rangle
\left( \phi _{47}^{+}+\phi _{47}^{-}\right) \\ 
\otimes \mid 0_1\rangle \left( \psi _{58}^{+}-\psi _{58}^{-}\right) -\mid
1_1\rangle \left( \phi _{58}^{+}+\phi _{58}^{-}\right) \\ 
\otimes \mid 0_1\rangle \left( \phi _{69}^{+}-\phi _{69}^{-}\right) -\mid
1_1\rangle \left( \psi _{69}^{+}+\psi _{69}^{-}\right)
\end{array}
\right] +\cdots \\
&&+\alpha _{1_71_81_9}{\cal R}\left[ 
\begin{array}{c}
\mid 0_1\rangle \left( \phi _{47}^{+}-\phi _{47}^{-}\right) -\mid 1_1\rangle
\left( \psi _{47}^{+}+\psi _{47}^{-}\right) , \\ 
\mid 0_1\rangle \left( \phi _{58}^{+}-\phi _{58}^{-}\right) -\mid 1_1\rangle
\left( \psi _{58}^{+}+\psi _{58}^{-}\right) , \\ 
\mid 0_1\rangle \left( \phi _{69}^{+}-\phi _{69}^{-}\right) -\mid 1_1\rangle
\left( \psi _{69}^{+}+\psi _{69}^{-}\right)
\end{array}
\right]  \nonumber
\end{eqnarray}
Make carefully collocation by the definition 2, we find 
\begin{equation}
\mid \Psi _{123456789}\rangle =\frac 18\sum_{R,S,T=1}^4\mid \varphi _{\left(
rst\right) }^{\prime }\rangle \otimes {\Bbb B}_{4,5,6,7,8,9}^{\left(
R,S,T\right) },\;\mid \varphi _{\left( R,S,T\right) }^{\prime }\rangle
=U_{RST}\mid \varphi _{123}\rangle
\end{equation}
where $\mid \varphi _{123}\rangle \equiv $the result of substitution $\mid
i_{r+6}$ $\rangle \longrightarrow \mid i_r$ $\rangle $ $\left( r=1,2,3\text{
and }i=0,1\right) $ from $\mid \varphi _{789}\rangle $ (notice that $\alpha
_{RST}$ does not change), i.e. $\mid \varphi _{123}\rangle =\sum_{i,j,k=0,1}$
$\alpha _{i_7j_8k_9}\mid i_1j_2k_3\rangle ,$ and $U_{RST}\left(
R,S,T=1,2,3,4\right) $ are sixty-four 8$\times 8$ unitary matrices 
\begin{eqnarray}
U_{RST} &=&U_R\otimes U_S\otimes U_T,\;  \nonumber \\
U_1 &=&U_{\phi ^{+}}=i\sigma _y=\left[ 
\begin{array}{ll}
0 & 1 \\ 
-1 & 0
\end{array}
\right] ,U_2=U_{\phi ^{-}}=-\sigma _x=\left[ 
\begin{array}{ll}
0 & -1 \\ 
-1 & 0
\end{array}
\right] \\
U_3 &=&U_{\psi +}=\sigma _z=\left[ 
\begin{array}{ll}
1 & 0 \\ 
0 & -1
\end{array}
\right] ,U_4=U_{\psi ^{-}}=-\sigma _0=\left[ 
\begin{array}{ll}
-1 & 0 \\ 
0 & -1
\end{array}
\right]  \nonumber
\end{eqnarray}
$\sigma _0$ is the unit matrix, and $\sigma _x,\sigma _y,\sigma _z$ are the
ordinary Pauli matrices. It is very interesting that $U_R\left(
R=1,2,3,4\right) $ are just those unitary matrices appearing in the BBCJPW
scheme[2] when $\mid \psi _{12}^{-}\rangle $ is taken as channel. The rest
steps are standard, i. e. Alice, respectively, makes three Bell measurements
for particle pairs (4,7), (5,8) and (6,9), she will obtain one of $\left\{ 
{\Bbb B}_{456789}^{\left( R,S,T\right) }\right\} $ with equal probability $%
\frac 1{64},$ simultaneously Bob obtain a corresponding $\mid \varphi
_{\left( RST\right) }^{\prime }\rangle \mid .$ When Alice informs her
measurement to Bob by classical communication, then Bob at once knows the
correct result must be $\mid \varphi _{123}\rangle =U_{RST}^{-1}\mid \varphi
_{\left( RST\right) }^{\prime }\rangle =U_R^{-1}\otimes U_S^{-1}\otimes
U_T^{-1}\mid \varphi _{\left( RST\right) }^{\prime }\rangle $, etc., so the
teleportation of N-partite qubit entangled state $\mid \varphi _{789}\rangle 
$ from Alice to Bob is completed.

If we take other ${\Bbb B}_{456789}^{\left( R,S,T\right) }$ as the quantum
channels, the results are similar. It can be verified that the above method
can be directly generalized to arbitrary unknown N-partite qubit entangled
states, i.e. if we take a ${\Bbb B}_{1\cdots 2N}^{\left( R_1,\cdots
,R_N\right) }\left( R_1,\cdots ,R_N=1,\cdots ,N\right) $ as quantum channel,
and Alice holds an arbitrary unknown N-partite qubit entangled state $\mid
\varphi _{\left( 2N+1\right) \cdots 3N}\rangle =\sum_{i_{2N+1},\cdots
,i_{3N}=0,1}$ $\alpha _{i_{2N+1}\cdots i_{3N}}\mid i_{2N+1}\cdots
i_{3N}\rangle ,$ she makes N Bell measurements for particle pairs $%
(N+1,2N+1),\cdots ,\left( 2N,3N\right) ,$ then Alice must obtain one of $%
\left\{ {\Bbb B}_{N+1,\cdots ,3N}^{\left( S_1,\cdots ,S_N\right) }\right\} $
with equal probability 2$^{-2N},$ and Bob obtain a state $\mid \varphi
_{1\cdots N}^{\prime }\rangle =U_{S_1,\cdots ,S_N}^{\left( R_1,\cdots
,R_N\right) }\mid \varphi _{\left( R_1,\cdots ,R_N\right) }^{\prime }\rangle 
$, where $\mid \varphi _{\left( R_1,\cdots ,R_N\right) }^{\prime }\rangle
=\sum_{i_{2N+1},\cdots ,i_{3N}=0,1}$ $\alpha _{i_{2N+1}\cdots i_{3N}}\mid
i_1\cdots i_N\rangle $. Especially we find that if in BBCJPW scheme the
unitary matrix corresponding to channel $\mid \Omega _{k,N+k}^{\left(
R_k\right) }\rangle \left( R=1,2,3,4\right) $ is $U_{S_{R_k}}^{\left(
R_k\right) }$ (i.e. a unknown state $\mid \varphi _{2N+k}\rangle =\alpha
\mid 0_{2N+k}\rangle +\beta \mid 1_{2N+k}\rangle $ is in Alice, when Alice
makes joint measurement for particle pair $(N+k,2N+k)$, the total wave
function $\mid \Omega _{k,N+k}^{\left( R_k\right) }\rangle \mid \varphi
_{2N+k}\rangle $ collapses to one $\left( U_{S_{R_k}}^{\left( R_k\right)
}\mid \varphi _k^{\prime }\rangle \right) \otimes \mid \Omega _{2,3}^{\left(
S_{R_k}\right) }\rangle (\mid \varphi _k^{\prime }\rangle =\alpha \mid
0_{k1}\rangle +\beta \mid 1_k\rangle ,\;S_{R_k}=1,2,3,4)$ with equal
probability $\frac 14),$ then in our scheme {\em \ } 
\begin{equation}
U_{S_1,\cdots ,S_N}^{\left( R_1,\cdots ,R_N\right) }=\otimes
_{k=1}^NU_{S_{R_k}}^{\left( R_k\right) }
\end{equation}
holds and Bob knows the correct result must be $\mid \varphi _{1\cdots
N}\rangle =\left( \otimes _{k=1}^N\left( U_{S_{R_k}}^{\left( R_k\right)
}\right) ^{-1}\right) \mid \varphi _{R_1,\cdots ,R_N}^{\prime }\rangle $.
This conclusion tall us that {\bf the teleportation of arbitrary
multipartite qubit entanglement, in fact, is essentially determined by the
teleportation of every single unknown qubit state.} In fact, this is
necessary. In fact, the entanglement of the state must invariant in any
teleportation, this means that the above teleportation must become N
independent teleportation for a separable state $\mid \varphi _{\left(
2N+1\right) \cdots 3N}\rangle =\otimes _{s=1}^N\mid \varphi _{2N+s}\rangle ,$%
where every  $\mid \varphi _{2N+s}\rangle =\alpha _{2N+s}\mid
0_{2N+s}\rangle +\beta _{2N+s}\mid 1_{2N+s}\rangle \left( s=1,\cdots
,N\right) $ is a qubit pure-state$,$ i.e. Eq.(16) must hold for $\mid
\varphi _{\left( 2N+1\right) \cdots 3N}\rangle $\ (where $%
U_{S_{R_k}}^{\left( R_k\right) }$ correponds to $\mid \varphi _{2N+k}\rangle 
$). Therefore Eq. (16) generally hold, because every $U_{S_1,\cdots
,S_N}^{\left( R_1,\cdots ,R_N\right) }$ is only dependent on indices $\left(
R_1,\cdots ,R_N\right) ,$ $\left( S_1,\cdots ,S_N\right) $.

{\it Conclusion.} We give a general scheme of teleportation of arbitrary
multipartite qubit entanglement. The return operators and cross product
operators are convenient in this scheme. Every one of cross Bell bases can
be taken as the quantum channel to realize the teleportation of arbitrary
multipartite qubit entanglement, this teleportation is essentially
determined by the teleportation of every single unknown qubit state.

\

\end{document}